\documentclass[conference]{IEEEtran}
\usepackage{fancyhdr}
\fancypagestyle{plain}{
  \fancyhf{}
  \fancyhead[C]{Invited demo at IEEE Cloudnet 2022}     

}
\usepackage{blindtext}
\usepackage{multirow}
\usepackage{amsmath,amsfonts}
\usepackage{algorithmic}
\usepackage{graphicx}
\usepackage{textcomp}
\usepackage{xcolor}
\usepackage{tikz,pgfplots}

\usepackage{caption}
\usepackage{subcaption}

\usepackage{comment}
\usepackage{float}
\usepackage{url}
\usepackage{diagbox}

\begin{document}

\newcommand{\todo}[1]{\textcolor{red}{#1}} 
\newcommand{\mluqos}[0]{\texttt{MLU-QoS}}
\newcommand{\mlu}[0]{\texttt{MLU}}
\newcommand{\allTNs}[0]{\texttt{All-TNs}}

\title{Routing and QoS Policy Optimization in SD-WAN}

\author{

\IEEEauthorrefmark{1}Pham Tran Anh Quang,
\IEEEauthorrefmark{1}Xu Gong, 
\IEEEauthorrefmark{1}Chuang Liu, 
\IEEEauthorrefmark{1}Kun Li, 
\IEEEauthorrefmark{1}J\'er\'emie Leguay\\
\IEEEauthorrefmark{2}Xinhui Zhang,
\IEEEauthorrefmark{2}Yong Zhang, 
\IEEEauthorrefmark{2}Jiayan Li,
\IEEEauthorrefmark{2}Kai Ye
\\

\\
\IEEEauthorblockA{\IEEEauthorrefmark{1}\textit{Huawei Technologies Ltd.}, \IEEEauthorrefmark{2}\textit{Industrial and Commercial Bank of China }}
}

\maketitle
\thispagestyle{plain}
\pagestyle{plain}

\begin{abstract}
In modern SD-WAN networks, a global  controller continuously  optimizes application and user intents by selecting the proper routing policies for each application. Nevertheless, the competition between flows can still occur at each overlay link and it may degrade Quality of Service (QoS).
To mitigate performance  degradations in case of congestion, QoS policies can also be dynamically optimized to limit the rate of low-priority flows and share the bandwidth among all flows in a fair manner.
This demonstration presents a comprehensive control plane architecture to coordinate path selection and rate allocation, in order to meet application requirements while optimizing a global objective (e.g., low congestion, best quality, and minimum cost).  
\end{abstract}

\section{Introduction}
Financial companies, as other businesses, are transforming their WAN (Wide Area Network) to reduce costs while optimizing user experience. To do so, they are  migrating from fully dedicated private lines, using MSTP (Multi-service SDH-based Transport Platform) for instance, to a cost-efficient mix of multiple access networks using SD-WAN.   
In particular, they are smoothly upgrading their WAN
by 1) downsizing some of the existing MSTP lines to reduce cost and 2) combining them with cheaper and shared private leased lines based on MPLS-VPN (MV).
While MSTP lines can guarantee strict Service Level Agreements (SLA) requirements, the reduction of their capacity along with the use of shared private lines, offering weaker QoS guarantees, is calling for advanced traffic engineering mechanisms based on dynamic load balancing and queuing. Indeed, the main challenge is to intelligently use all the available capacity to meet applications requirements and optimize user experience.

\begin{figure}[t]
    \centering
    \begin{subfigure}[b]{0.8\columnwidth}
        \includegraphics[width=\textwidth]{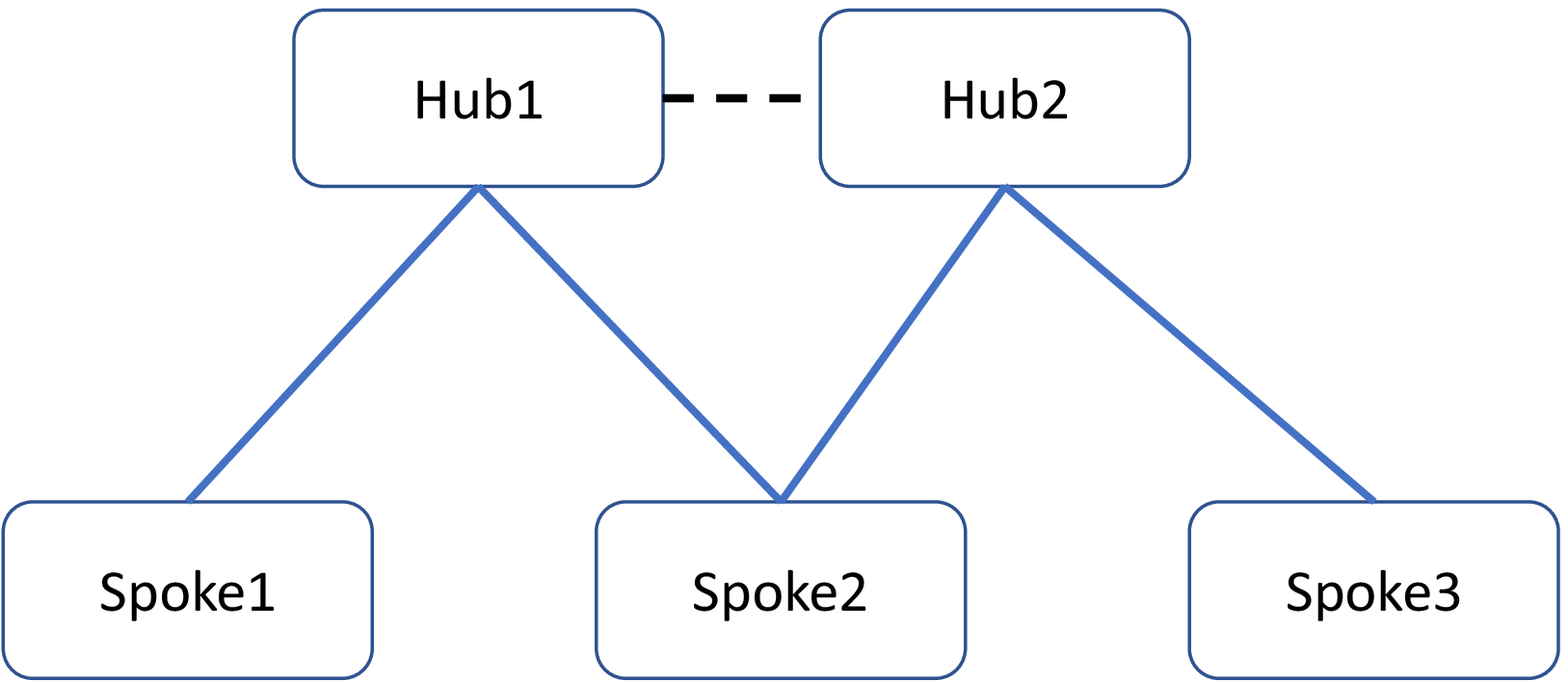}
        \caption{Access network with MSTP only}
        \label{fig:MSTP}
    \end{subfigure}
    \begin{subfigure}[b]{0.8\columnwidth}
        \includegraphics[width=\textwidth]{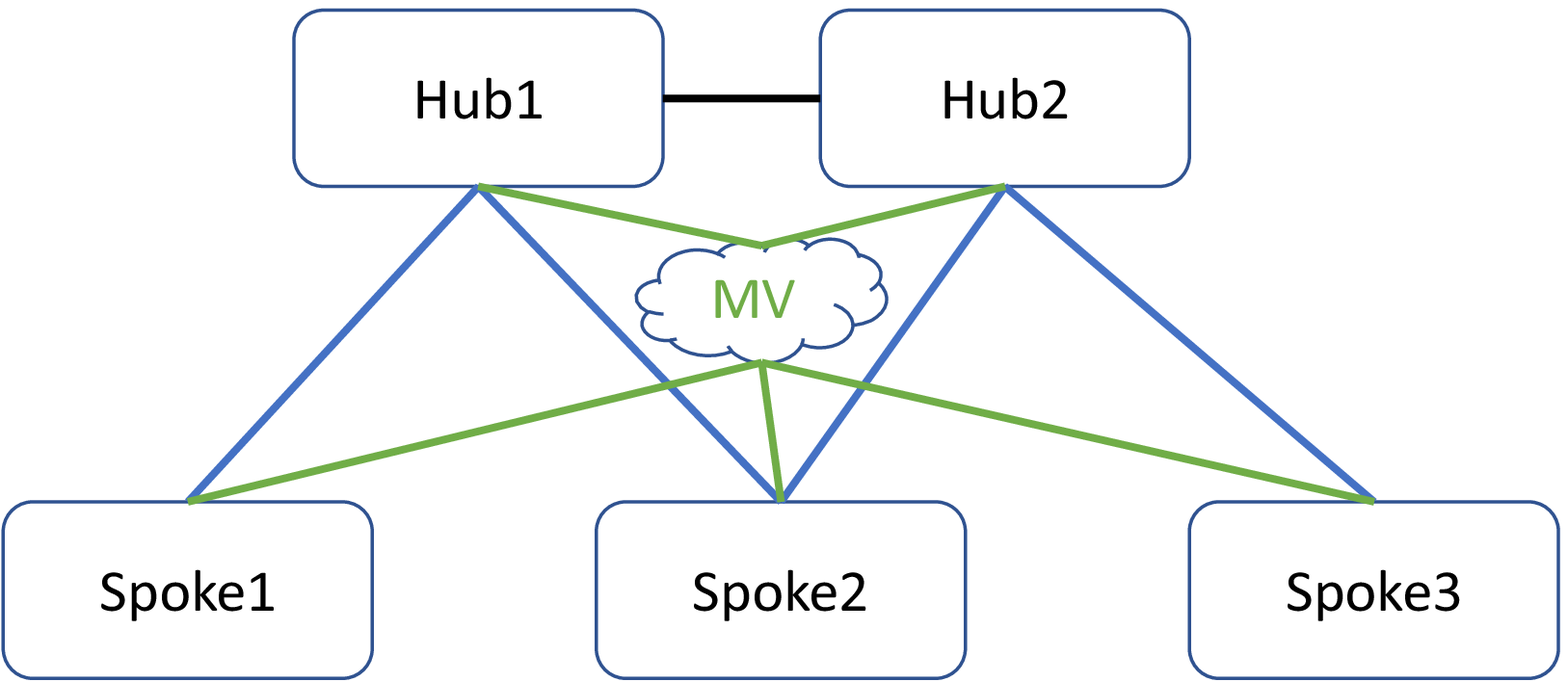}
        \caption{SD-WAN network combining MSTP, MV, and MAN}
       \label{fig:combined}
    \end{subfigure}
    \caption{Network transformation scenario from a) to b).}
    \label{figtopo}
\end{figure}

This demonstration will consider a use case from the Industrial and Commercial Bank of China (ICBC), illustrated in Fig.~\ref{figtopo}, where $2$ hubs (regional branches) are connected to 3 spokes (other sites except regional branches). In the original setup, MSTP lines at $18$ Mpbs connect hubs to spokes (capacities have been scaled down to keep simulations tractable but proportions are kept). The Metropolitan Area Network (MAN) between hubs is not used as all flows can only be routed over a single path (flow-level multi-path traffic steering is not available). After migration to SD-WAN, existing MSTP lines are downsized by $66\%$ to save cost and they are combined with other access networks: 1) existing hub-to-hub connectivity can now be used as one routing option between hubs and spokes, and 2) shared MV lines at $12$ Mbps are subscribed. 
The total cost after network transformation is reduced by $26$\% approximately.

We previously demonstrated a policy optimization algorithm for Smart Policy Routing (SPR)~\cite{huawei-SPR} to optimize global intents (e.g., minimum cost, best delay), while meeting QoS requirements of applications~\cite{anh2021intent, quangintent}. This demonstration is a further improvement with a joint optimization of SPR and QoS policies. Indeed, while load balancing greatly improves performance, dynamic QoS policies can further improve performance in case of congestion or tight SLA requirements. 
In this demonstration, the network controller periodically decides the set of overlay links that each application, also called \textit{flow group}, is allowed to use (i.e., SPR policies), and the QoS policy that is applied to each flow group on each overlay link. 

A number of Active Queue Management (AQM) techniques have been proposed to help sustain delay and throughput requirements~\cite{adams2012active}. Most of them uses implicit (packet dropping) or explicit signals (ECN marking) to control sources sharing bottleneck link. They do not aim at tuning queuing parameters and optimizing performance globally. Other works on Adaptive Weighted Fair Queuing (AWFQ) have proposed the dynamic adaptation of scheduling parameters. However, they consider the control of only one device (local approach, no coordination with among devices)~\cite{frantti2009embedded}, a distributed approach using bandwidth pricing to coordinate multiple devices~\cite{sayenko2006comparison} or a distributed protocol to trigger queue adjustments~\cite{hussain2003agent}. In the later solution, agents at the destination informs about delay violations so that upstream agents adjust their weights.

Our solution goes beyond state-of-the-art, as it defines a global framework to optimize routing and queuing parameters at the centralized controller to coordinate all edge devices in the context of overlay networks. Existing solutions do not consider a global optimization and do not perform a joint optimization of routing and QoS. 

\begin{figure}[t]
    \centering
    \includegraphics[width=\columnwidth]{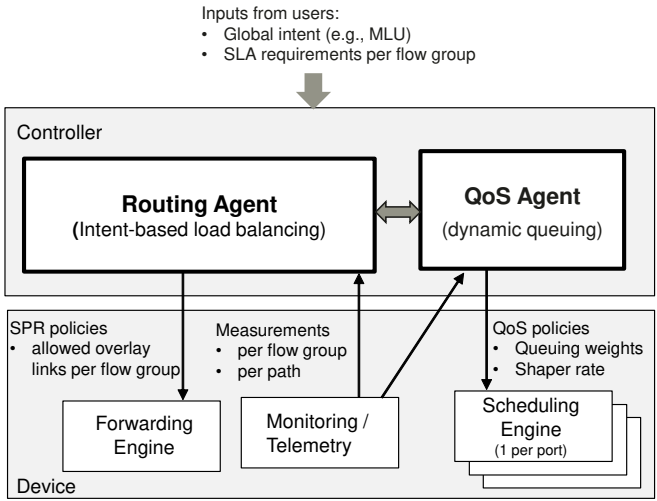}
    \caption{System architecture with the  controller and routers.}
    \label{fig:arch}
\end{figure}

\section{Joint policy optimization}
As depicted in Fig.~\ref{fig:arch}, a centralized controller can modify periodically routing policies and queuing parameters with corresponding agents.  In this demonstration, we consider the case where these agents operate independently and in a time-triggered manner. 
Routing policies are updated at a slow pace to leave some time for the load balancing to converge, while queuing policies are updated more frequently as they immediately impact performance. 

\begin{figure}[t]
    \centering
    \includegraphics[width=\columnwidth]{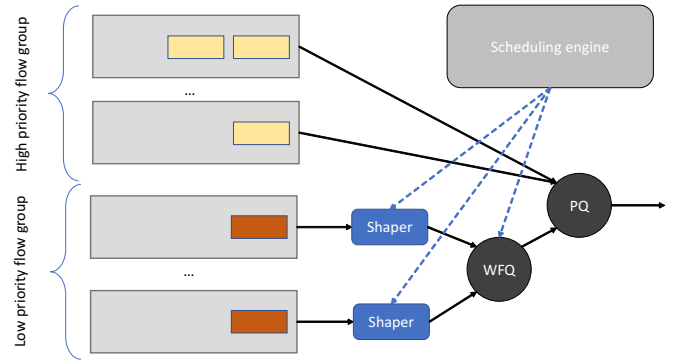}
    \caption{Class-Based Queuing (CBQ) scheduler}
    \label{fig:cbq}
\end{figure}

As presented in Fig.~\ref{fig:cbq}, we consider a hierarchical QoS scheduler based on Class-Based Queuing (CBQ) that handles high priority flows with Priority Queuing (PQ) and low priority flows with Weighted Fair Queuing (WFQ). A CBQ scheduler is deployed at each port of each edge router (at hubs and spokes) while the other devices in the middle (in the underlay) operate using a single First In First Out (FIFO) queue at each port. The bandwidth allocation for each of the low priority flow groups is controlled thanks to shapers (i.e., maximum rate) and scheduling weights (minimum rate).

\textbf{Intent-based routing and smart queuing.}
Ingress routers are configured with SPR and QoS policies for each flow group. The SPR policy~\cite{huawei-SPR} contains the set of allowed overlay links that a flow group can use. The QoS policy defines queuing parameters, e.g. queue weights and shaper rates. At each ingress device, the sub-set of active paths over which each flow group is load balanced is determined using link-level measurements (i.e., delay, jitter, loss). In case of congestion, the performance of high-priority flows can degrade and bandwidth starvation can occur for low-priority flows. Therefore, the dynamic tuning of QoS policies aims at  protecting high-priority flow groups and  while ensuring fairness in performance degradation for low-priority flow groups. In this demonstration, we present an automatic configuration scheme for both SPR and QoS policies to enhance SLA satisfaction for all flow groups.

\textbf{Problem formulations.}
We consider a set of overlay links $E$ of capacity $C_e$ and a set of flow groups $K$. For each flow group $k\in K$, the measured traffic demand is denoted $b^k$ and the set of all possible overlay links is given by $E_k\subseteq E$ (e.g., outgoing links of ingress routers). The delay requirement of flow group $k$ is $D_k$.

\textbf{1) SPR policy optimization problem.} The following model optimizes a global intent (e.g. minimize congestion or maximize quality) while meeting SLAs  (see~\cite{quangintent} for model details).

\begin{alignat}{3}
\min \quad & \sum_{k \in K}\sum_{e \in E_k} \bigl[ \sum_{i \in I_e} MLU \text{ (congestion) or } f_e^k(x) \text{ (quality)} \bigr] \nonumber\\
\textrm{s.t.}  \quad & MLU\geq \sum_{i \in I_e} LU_e^i = \sum_{k\in K_e} b_k x_e^k \leq C_e ,  \quad \forall e\in E, \label{capacity_constraints}\\
  & f_e^k(x) \leq D_k \quad \forall k\in K,\forall e\in E_k,   \label{delay_constraints}\\
  & \sum_{e\in E_k} x_e^k =1,\quad \forall k\in K   \label{convexity_constraints}
\end{alignat}
Once SPR policies have been optimized through a \textit{slow control loop},
the traffic for flow group $k$ over overlay link $e$ is expected to be $d^k_e = b^k x_e^k$. To further optimize bandwidth sharing at each port, QoS policies are optimized over a \textit{faster control loop} thanks to the following model.

\textbf{2) QoS policy optimization problem.}

\begin{alignat}{3}
\min \quad & \sum_{e} \left( -d^k_e \log{z^k_e} + \sum_{k:d^k_e >0} \left( \alpha \frac{y^k_e}{D_k}+M^k h^k_e\right) \right) \nonumber\\
\textrm{s.t.} \quad & \sum_{k} z^k_e \leq C_e ,  \quad \forall e\in E, \label{capacity_constraints}\\
  & f^k_e(z) \leq D_k + y_e^k \quad \forall k\in K,\forall e\in E_k,   \label{delay_constraints}\\
  & z^k_e+h^k_e \geq d^k_e,\quad \forall k\in K,\forall e\in E   \label{demand_constraints}
\end{alignat}
The rate allocation violation $h_e^k$ identifies the gap between the demand and the allocated rate. The SLA violation $y_e^k$ is the gap between SLA requirements (e.g., end-to-end delay) and  prediction. The SLA prediction $f_e^k(z)$ is a function of the rate allocation that could be derived from queuing theory or a data-driven model. The objective of the  optimization problem is to (i) satisfy the traffic demand ($M^k$ is large to prioritize this objective), (ii) ensure fairness, and (iii) minimize SLA violations. We introduce $\alpha$ to trade-off fairness with SLA violations. In this demonstration, we select $\alpha=0.01$ to give more priority to the bandwidth fairness. The output is the rate allocation for each flow group, $z^k_e$, which is used to tune queuing weights and shaping.

\begin{figure}[t]
     \centering
     \includegraphics[width=\columnwidth]{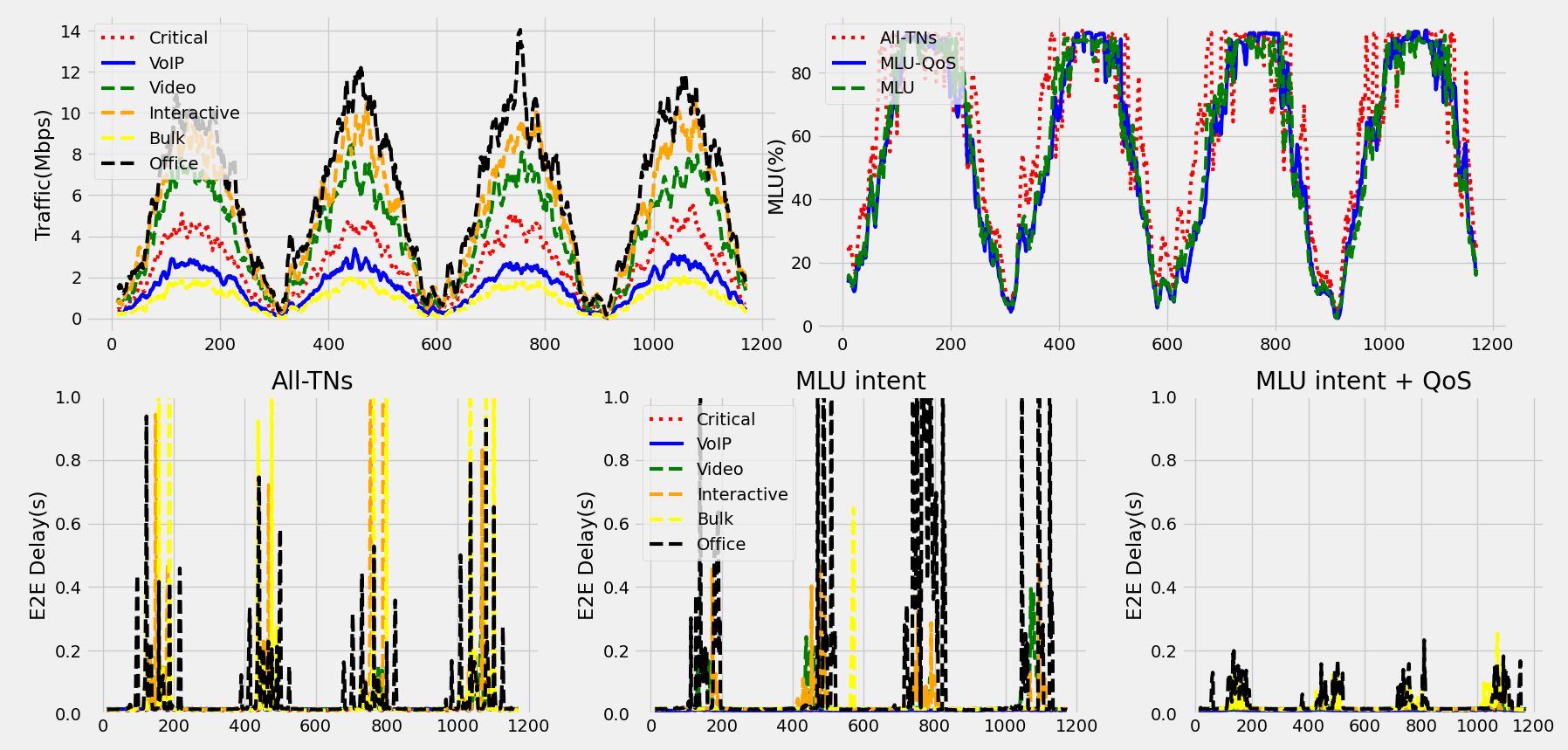}
     \caption{Visualization of the total traffic and the end-to-end delay for each application, and the MLU.}
     \label{fig:gui}
\end{figure}

SPR and QoS optimization problems can be solved using a solver (e.g. SCIP) and 
heuristic algorithms like the local search algorithm are able to optimize SPR policies efficiently.
\begin{table}[b]
\centering
\begin{tabular}{|p{3cm}|p{2.5cm}|p{2.5cm}|p{2.5cm}|p{2.5cm}|}
 \hline 
  \diagbox[width=\dimexpr \textwidth/6+2\tabcolsep\relax, height=0.7cm]{App}{Policy} &\allTNs(a)& \allTNs(b)&\mlu(b)&\mluqos(b) \\
 \hline
 &\multicolumn{4}{c|}{APP SLA satisfaction rate (\%) / Avg E2E delay (ms) / $95^{th}$ E2E delay (ms)} \\
 \hline
Critical & 100 / 10.6 / 10.8 & 78 / 27.9 / 127 & 100 / 11.0 / \ 11.3 & 1.00 / 11.7 / \ 13.3\\
\hline
VoIP & 100 / 10.6 / 10.8 & 84 / 27.3 / 105 & \ 95 / 15.0 / \ 17.9 & 100 / 14.4 / \ 17.4 \\
\hline
Video & 100 / 11.0 / 11.2 & 79 / 56.2 / 249 & \ 93 / 28.8 / 157.3 & 100 / 16.6 / \ 18.3 \\
\hline
Interactive & 100 / 11.1 / 11.5 & 77 / 65.4 / 262 & \ 95 / 24.2 / \ 43.5 & 100 / 17.2 / \ 19.8 \\
\hline
Bulk & 100 / 11.2 / 12.0 & 92 / 45.8 / 192 & 100 / 17.7 / \ 18.9 & \ 98 / 23.6 / \ 69.4 \\
\hline
Office & 100 / 12.0 / 14.0 & 85 / 64.1 / 291 & \ 91 / 50.8 / 243.0 & \ 98 / 33.2 / 121.0 \\
\hline
\end{tabular}
\caption{Performance for all policy optimization scenarios}
\label{tab:delay}
\end{table}

\section{Demonstration}
We use NS3~\cite{Riley2010} and
applications are classified into $6$ flow groups with increasing end-to-end delay requirements and priority: Critical~($15$ms), VoIP~($20$ms), Video~($40$ms), Interactive~($30$ms), Bulk~($100$ms) and Office~($150$ms). 
Each spoke receives $6$ flow groups from servers located either at hub 1 or 2. The traffic demand is the same in scenarios of Fig.~\ref{figtopo} (a\&b). 
The transport layer is TCP. The microflow inter-arrival time in each flow group varies to generate diurnal traffic patterns. 
The propagation delay of MSTP and MV links are $10$ms and $15$ms, respectively. A Hierarchical QoS scheduler is deployed to prioritize traffic. The $4$ highest priority applications can preempt other applications (strict priorities). Bandwidth sharing among the $2$ lowest priority flow groups is ensured using WFQ and their individual rate can be shaped. Packet priorities are marked using DSCP. Link-level measurements are collected every $1$s. SPR and QoS policies are updated by the controller every $50$s and $10$s, respectively (periods may depend on traffic variations).

Table~\ref{tab:delay} provides End-to-End (E2E) delay (average and $95^{th}$ percentile) and SLA satisfaction rate (\% of epochs SLA requirement is met) for each flow group. Three configurations of SPR and QoS policies are considered: (i) \allTNs~where all links can be used by ingress devices (SPR policies are not optimized by the controller) (ii) \mlu~ where SPR policies are optimized to minimize congestion (i.e., Maximum Link Utilization) but  fixed queuing parameters are used, and (iii) \mluqos~where SPR and QoS
policies are jointly optimized. 
For \allTNs, the two scenarios of Figs.~\ref{figtopo}a\&b are compared, denoted as \allTNs(a) and \allTNs(b). The first scenario has higher MSTP capacities but higher deployment cost. 
As a consequence, it can satisfy all SLA requirements. \mluqos~has the best performance for the SD-WAN network configuration. It can guarantee SLAs in more than $95\%$ of times. Without QoS policy optimization,  low-priority flows experience bandwidth starvation, which causes SLA violations. In addition,  high-priority flows group are not protected against low-priority flows.

Fig.~\ref{fig:gui} presents the graphical interface of our demo: a dashboard to visualize for all policy optimization strategies the total traffic and the end-to-end delay, and the congestion (i.e.,  MLU). The video is here: \url{https://tinyurl.com/3atckfpf}
\bibliographystyle{IEEEtran}
\bibliography{reference}

\end{document}